\documentclass[conference]{IEEEtran}
\usepackage{color,graphicx}
\usepackage{url}
\usepackage{amsmath}
\usepackage{algorithmic}
\usepackage{epstopdf}
\usepackage{epsfig}
\usepackage{todonotes}

\makeatletter

\ifCLASSINFOpdf
\else
\fi

\makeatletter

\newcommand{\Rmnum}[1]{\expandafter\@slowromancap\romannumeral #1@}
\makeatother

\begin{document}
\title{Machine Learning based detection of
multiple \mbox{Wi-Fi} BSSs for LTE-U CSAT}
\author{Vanlin Sathya$^\dag\text{*}$, Adam Dziedzic$^\dag\text{*}$, Monisha Ghosh$^\dag$, and Sanjay Krishnan$^\dag$
\IEEEauthorblockN{}
\IEEEauthorblockA{$^\dag$University of Chicago, Illinois, USA.\\
{Email: vanlin@uchicago.edu, ady@uchicago.edu, monisha@uchicago.edu, skr@uchicago.edu.}}
}
\maketitle

\footnote{*Equal contribution.}
\begin{abstract}
According to the LTE-U Forum specification, a LTE-U base-station (BS) reduces its duty cycle from 50\% to 33\% when it senses an increase in the number of co-channel \mbox{Wi-Fi} basic service sets (BSSs) from one to two. The detection of the number of \mbox{Wi-Fi} BSSs that are operating on the channel in real-time, without decoding the \mbox{Wi-Fi} packets, still remains a challenge. In this paper, we present a novel machine learning (ML) approach that solves the problem by using energy values observed during \mbox{LTE-U} OFF duration. Observing the energy values (at \mbox{LTE-U} BS OFF time) is a much simpler operation than decoding the entire \mbox{Wi-Fi} packets. In this work, we implement and validate the proposed ML based approach in real-time experiments, and demonstrate that there are two distinct patterns between one and two \mbox{Wi-Fi} APs. This approach delivers an accuracy close to 100\% compared to auto-correlation (AC) and energy detection (ED) approaches.
\end{abstract}

\section{Introduction}\label{sec:introduction}
The inherent challenge of \mbox{Wi-Fi} and  Long Term Evolution (LTE) coexisting fairly in the unlicensed bands at 5 GHz has been addressed by recent standardization efforts: LTE-LAA developed by 3GPP~\cite{3gpp} and LTE-U developed by the LTE-U Forum~\cite{forum}. These two standardization entities differ in the way coexistence is implemented. LTE-LAA~\cite{TCCN} uses a mechanism similar to CSMA/CA as used in \mbox{Wi-Fi}, also known as Listen Before Talk (LBT), while LTE-U leverages a duty cycle approach (\emph{i.e.,} repeating ON and OFF intervals) and an adaptation technique called Carrier Sense Adaptive Transmission (CSAT). During an ON period, the LTE-U BS transmits its data normally. In the OFF period, it observes the channel and dynamically adjusts its duty cycle based on the number of detected \mbox{Wi-Fi} basic service sets (BSSs) or access points (APs). The detection method of \mbox{Wi-Fi} BSSs is arguably still a point of contention. Table~\ref{table:csat} shows different types of possible CSAT approaches: directly decoding the \mbox{Wi-Fi} MAC header of \mbox{Wi-Fi} BSSs or spectrum sensing using energy detection (ED), auto-correlation (AC), or machine learning (ML) models. Each method has its own pros and cons as listed in the Table~\ref{table:csat}. In our previous work  \cite{sathya2018energy}\cite{AC}~\footnote{The latest version can be found here: http://bit.ly/2LDVWWo}, we studied ED and AC based detection of \mbox{Wi-Fi} APs, and demonstrated algorithms that performed reasonably well under different scenarios. 

\par Of late, machine learning (ML) approaches are beginning to be used in wireless networks to solve problems such as agile management of network resources using real-time analytics based on data. ML models enable us to replace heuristics with more robust and general alternatives. In this paper, we collect the \mbox{Wi-Fi} AP energy values during LTE-U OFF duration and use the data to train different ML models~\cite{DBLP:journals/corr/abs-1803-04311}. We also apply the models in an online experiment to detect the number of \mbox{Wi-Fi} APs. Finally, we demonstrate significant improvement in the performance of the ML approach as compared to the ED and AC detectors.   

\begin{table}
\caption{Different Types of CSAT.
}
\centering	
\begin{tabular}{|p{1.5cm}| p{2cm}| p{1.8cm}| p{1.8cm} |}
\hline
\bfseries
CSAT Types &\bfseries Method &\bfseries Pros &\bfseries Cons \\ 
\hline
Header Decoding (HD) & Decodes the \mbox{Wi-Fi} MAC header at the \mbox{LTE-U} BS & 100\% accurate & Additional Complexity~\cite{Chai:2016:LUS:2973750.2973781}, high cost\\
\hline
Energy Detection (ED) & Based on the change in the \textit{energy level} of the air medium & 
Low-cost, low-complexity & Low-accuracy
\cite{sathya2018energy}\\
\hline
Auto-correlation (AC) & LTE-U BS performs correlation on the \mbox{Wi-Fi} L-STF symbol in the preamble & Low-cost, low-complexity & Medium accuracy (more accurate than ED)~\cite{AC}  \\
\hline
Machine Learning (ML) & Train the model based on energy values on the channel & Much more accurate than ED and AC methods & Requires gathering data and training models\\
\hline
\end{tabular}
\label{table:csat}
\end{table}

\begin{figure}[htb!]
\begin{center}
\includegraphics[height=3cm,width=9cm]{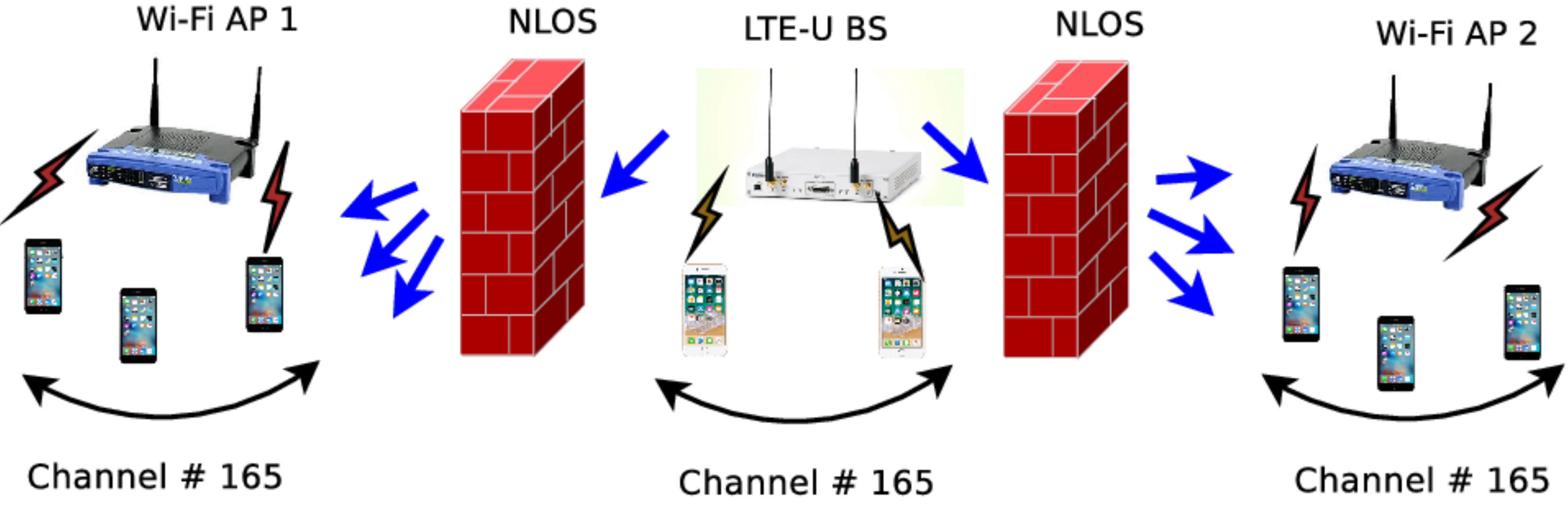}
\caption{LTE \mbox{Wi-Fi} Co-existence Deployment Setup.}
\label{expp}
\end{center}
\end{figure}

\par Fig.~\ref{expp} illustrates an example LTE-U/\mbox{Wi-Fi} coexistence scenario, where two \mbox{Wi-Fi} APs and one \mbox{LTE-U} BS are operating on the same channel, with multiple clients associated with each AP and BS. According to 3GPP, it is expected that the \mbox{LTE-U} BS will adjust its duty cycle from 33\% to 50\% when one of the APs is turned off, and vice versa. Hence, a robust and accurate detection method is needed to guarantee the efficient adaptation of the duty cycle. We aim to exploit the collected energy level data by using the ML approach at the LTE-U BS to infer the presence of one or two \mbox{Wi-Fi} BSSs and make the decision to adapt the duty cycle appropriately. In order to accomplish this, we create realistic experimental scenarios using a National Instruments (NI) USRP RIO board with a LTE-U module, two Netgear \mbox{Wi-Fi} APs, and two Wi-Fi clients. 

The rest of the paper is organized as follows. Section~\ref{sec:related-work} presents a brief overview of existing literature on \mbox{Wi-Fi} LTE coexistence. Section~\ref{sec:ac-setup} describes the experimental measurement set-up used to gather statistics on the energy values in the presence of one or two \mbox{Wi-Fi} APs which are then used to develop the ML adaptation algorithm described in Section~\ref{sec:ML}. The experimental results are presented in Section~\ref{sec:experimental-results}. Finally, Section~\ref{sec:conclusion} concludes the paper.

\section{Related Work}\label{sec:related-work}
The coexistence of LTE and \mbox{Wi-Fi} in the unlicensed spectrum gives rise to several challenges in terms of \mbox{Wi-Fi} client association, interference management, scheduling/resource allocation, fair coexistence, imperfect carrier sensing, etc. There has been a significant amount of research, both from academia and  industry on LTE/\mbox{Wi-Fi} coexistence. This is mainly driven by the strong intention both from 3GPP and LTE-U forum to implement the technology as soon as possible. Both  License Assisted Access (LAA)/\mbox{Wi-Fi} and LTE-U/\mbox{Wi-Fi} coexistence scenarios and throughput fairness have been well studied as a function of detection threshold and duty-cycle \cite{Chai:2016:LUS:2973750.2973781,cano2016unlicensed,chen2016optimizing}. However, the energy based and auto-correlation based techniques proposed in the existing literature are still under-utilized in the area of spectrum sensing for LTE-U/\mbox{Wi-Fi} coexistence. 
In our recent work,~\cite{sathya2018energy} and \cite{WCNC,ICC}, we performed rigorous theoretical and experimental analyses of the performance of an energy-based CSAT.
We proposed an algorithm that can adjust the duty cycle of LTE-U based on the presence of \mbox{Wi-Fi} APs inferred by the detected energy in the medium. We believe that this is the first work that proved the feasibility of stand-alone energy detection, without the need of packet decoding. We are able to reliably  distinguish the presence between one or two \mbox{Wi-Fi} APs, using a threshold of -42 dBm which produced a successful detection probability $P_D$ of greater than 80\%  and false positive probability $P_{FA}$ (false alarm) of less the 5\%. To further improve the performance of energy based approach in $P_D$ and $P_{FA}$, we propose a novel algorithm that solves the problem by using an auto correlation (AC) function~\cite{AC} on the \mbox{Wi-Fi} preamble and setting appropriate detection thresholds to infer the number of \mbox{Wi-Fi} BSSs operating on the channel. Performing auto-correlation on the \mbox{Wi-Fi} preamble is more accurate than the energy-based approach. We show that using an AC threshold of $N_E$ = 0.8, we can achieve a probability of detection ($P_D$) of 0.9 with a probability of false alarm ($P_{FA}$) of less than 0.02. In this paper, we aim to further improve the performance of AC detection by introducing an alternative efficient approach \emph{i.e.,} 
\textit{ML based decision} to distinguish between one and two \mbox{Wi-Fi} BSSs on the same channel.

\begin{figure}[htb!]
\begin{center}
\includegraphics[totalheight=5cm,width=8.2cm]{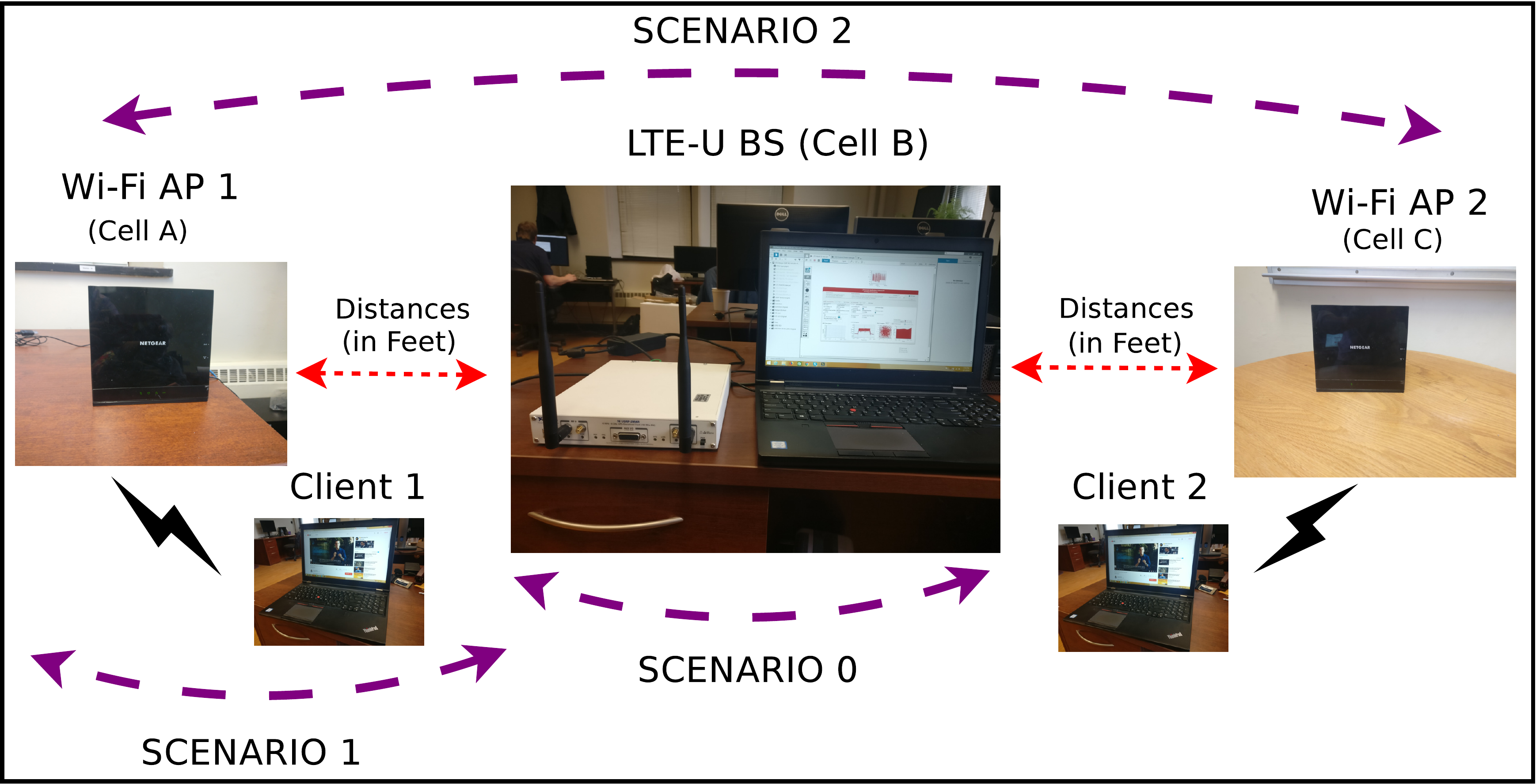}
\caption{LTE \mbox{Wi-Fi} Co-existence Experimental Setup.}
\label{exp}
\end{center}
\end{figure}

\begin{table}[htb!]
\caption{Experimental Set-up Parameters}
\centering
\begin{tabular}{|p{4cm}| p{4cm}|}
\hline\bfseries
Parameter&\bfseries Value \\ [0.4ex]
\hline
Available Spectrum and Frequency  & 20 MHz and 5.825 GHz \\
\hline
Maximum Tx power for both LTE and \mbox{Wi-Fi} & 23 dBm \\ 
\hline
Wi-Fi sensing protocol & CSMA/CA \\
\hline
Traffic & Full Buffer (Saturation Case) \\
\hline
Wi-Fi \& LTE-U Antenna Type & MIMO \& SISO\\
\hline
LTE-U data and control channel & PDSCH and PDCCH \\
\hline
\end{tabular}
\label{sim}
\end{table}
\section{Experimental Setup for Machine Learning Based Detection}\label{sec:ac-setup}
We set up our experimental test-bed according to Fig.~\ref{exp} and the experiment parameters are described in Table~\ref{sim}. An experiment is set up to evaluate the duty cycle adaptation algorithm performance considering LTE-U and \mbox{Wi-Fi} coexistence on the same unlicensed 20 MHz channel in the 5 GHz band. LTE-U utilizes the unlicensed spectrum only for the downlink and all uplink transmissions use the licensed spectrum. The LTE-U BS operates at maximum power by enabling all its resource blocks with the highest modulation coding scheme (i.e., 64-QAM). There are three cells, with two cells (A and C) acting as \mbox{Wi-Fi} cells and one cell (\emph{i.e.,} Cell B) acting as the LTE-U BS. We also ensure that there is no additional interference from other \mbox{Wi-Fi} APs on a particular channel. Each \mbox{Wi-Fi} cell consists of 1 AP and 1 client, and each AP transmits full buffer downlink data and beacon frames, with occasional probe responses if any nearby \mbox{Wi-Fi} clients transmit probe requests. An NI USRP platform is configured as the LTE-U BS and operates at 50\% duty cycle during the experiment. During the LTE-U OFF duration, it listens to the configured co-channel for signals and measures its indicator \textit{RF power}. The \textit{RF power} function is configured in the LTE block control module of the NI LTE application framework, and it outputs energy value as defined above. Using the energy measurement, we identify the number of \mbox{Wi-Fi} APs in the following scenarios: 
\begin{itemize}
\item \textbf{Scenario 0:} Measure the energy at the LTE-U BS during the
LTE-U OFF period when no \mbox{Wi-Fi} APs are deployed.
\item \textbf{Scenario 1:} Measure the energy at the LTE-U BS during the
LTE-U OFF period when only one \mbox{Wi-Fi} AP (\emph{i.e.,} Cell
A) and LTE-U (i.e., Cell B) is deployed with full buffer downlink transmission. 
\item \textbf{Scenario 2:} Measure the energy at the LTE-U BS during the
LTE-U OFF period when two \mbox{Wi-Fi} APs (i.e., Cell A \& Cell C) and LTE-U (i.e., Cell B) are deployed with full
buffer downlink transmission.
\end{itemize}
In each scenario, Cell B measures the energy values during the LTE-U OFF period, while other cells are transmitting full buffer downlink transmission. Also, these scenarios are carried out at different distances in both LOS and NLOS environment. Similar to our previous work~\cite{sathya2018energy, AC}, we focus only on Scenario 1 and 2 (i.e., 1 and 2 Wi-Fi). The Scenario 0 (i.e., no Wi-Fi) can be easily detected when there is a change in the energy values~\cite{ICC}.

\section{Machine Learning Approach}\label{sec:ML}

ML models enable us to replace heuristics with more robust and general alternatives. For the problem of distinguishing between zero, one, two \mbox{Wi-Fi} BSSs, we train a model to detect a pattern in the signals instead of finding a specific energy threshold in a heuristic manner. The state-of-the-art ML models leverage heavily the unprecedented performance of neural networks models that are able to surpass human performance on many tasks, for example, image recognition~\cite{SurpassHuman}, and help us answer complex queries on videos~\cite{DeepLens}. This efficiency is a result of large amounts of data that can be collected and labeled as well as usage of highly parallel hardware such as GPUs or TPUs~\cite{TPUshort,cuDNN}. In the work described in this paper, we train our neural network models on NVidia GPUs and collect enough data samples that enable our models to achieve high accuracy. Our major task is a classification problem to distinguish between zero, one, two \mbox{Wi-Fi} BSSs.

We consider machine learning models that take time-series data of width $w$ as input, giving an example space of $\mathcal{X} \in \mathcal{R}^{w}$, where $\mathcal{R}$ denotes the real numbers. Our discrete label space of $k$ classes is represented as $\mathcal{Y} \in \{0,1\}^k$. For example, $k=3$ classes, enables us to distinguish between 0, 1, and 2 \mbox{Wi-Fi}s. Machine learning models represent parametrized functions (by a weight vector $\theta$) between the example and label spaces $f(x;\theta): \mathcal{X} \mapsto \mathcal{Y}$. The weight vector $\theta$ is iteratively updated during the training process until the convergence of the train accuracy or train loss (usually determined by very small changes to the values despite further training), and then the final state of $\theta$ is used for testing and real-time inference.

\subsection{Data preparation}
The training and testing data is collected over an extended period of time; a single case (a single number of \mbox{Wi-Fi} AP) takes about 8 hours. For ease of exposition, we consider the case with one and two \mbox{Wi-Fi} APs. We collect data for each \mbox{Wi-Fi} AP independently and store the two datasets in separate files. Each file contains more than 2.5 million values and the total raw data size in CSV format is of about 60 MB. Each file is treated as time-series data with a sequence of values that are first divided into chunks. We overlap the time-series chunks arbitrarily by three-fourths of their widths $w$. For example, for chunks of width $w=128$, the first chunk starts at index 0, the second chunk is formed starting from index 32, the third chunk starts at index 64, and so on. This is part of our data augmentation and a soft guarantee that much fewer patterns are broken on the boundary of chunks. The width $w$ of the (time-series data) chunk acts as a parameter for our ML model. It denotes the number of samples that have to be provided to the model to perform the classification. The longer the time-series width $w$, the more data samples have to be collected during inference. The result is higher latency of the system, however, the more samples are gathered, the more accurate the predictions of the model. On the other hand, with smaller number of samples per chunk, the time to collect the samples is shorter, the inference is faster but of lower accuracy. We elaborate more on this topic in Section~\ref{sec:experimental-results}.

The collection of chunks are shuffled randomly. We divide the input data into training and test sets, each 50\% of the overall data size. The aforementioned shuffling ensures that we evenly distribute different types of patterns through the training and test sets so that the classification accuracy of both sets is comparable. Each of the training and test sets contain roughly the same number of chunks that represent one or two \mbox{Wi-Fi} BSSs. We enumerate classes from 0. For the case of 2 classes (either one or two \mbox{Wi-Fi}s), we denote by \textit{0} the class that represents a single \mbox{Wi-Fi} BSS and by \textit{1} the class that represents 2 \mbox{Wi-Fi} BSSs. Next, we compute the mean $\mu$ and standard deviation $\sigma$ only on the train set. We check for outliers and replace the values that are larger than $4\sigma$ with the $\mu$ value (e.g., there are only 4 such values in class \textit{1}). 

The data for the two classes have different ranges (from about -45.46 to -26.93 for class \textit{0}, and from about -52.02 to about -22.28 for class \textit{1}). Thus, we normalize the data $D$ in the standard way: $ND = \frac{(D - \mu)}{\sigma}$, where $ND$ is the normalized data output, $\mu$ and $\sigma$ are the mean and standard deviation values computed on the train data. We attach the appropriate label to each chunk of the data. The overall size of the data after the preparation to detect one or two \mbox{Wi-Fi} APs is of about 382 MB, where the \mbox{Wi-Fi} APs are on opposite sides of \mbox{LTE-U} BSS and placed at 6 feet distance from the \mbox{LTE-U} BSS). We collect data for many more scenarios and present them in Section~\ref{sec:experimental-results}. The final size of the collected data is 3.4 GB. 

For training, we do not insert values from different numbers of \mbox{Wi-Fi} APs into a single chunk. The received signal in the \mbox{LTE-U} BSS has higher energy on average for more \mbox{Wi-Fi} APs, thus there are differences in the mean values for each dataset. 
Our data preparation script handles many possible numbers of \mbox{Wi-Fi} APs and generates the data in the format that can be used for model training and inference (we follow the format for datasets from the UCR archive). In the future, we plan on gathering additional data samples for more Wi-Fi APs and make the dataset more challenging for classification.

\subsection{Neural network models: FC, VGG and FCN}
Our data is treated as a uni-variate time-series for each chunk. There are many different models proposed for the standard time-series benchmark~\cite{UCRArchive2018}. 

First, we test \textit{fully connected (FC)} neural networks. For simple architectures with two linear layers followed by the ReLU non-linearity the maximum accuracy achieved is about 90\%. More linear layers, or using other non-linearities (e.g. sigmoid) and weight decays do not help to increase the accuracy of the model significantly. Thus, next we extract more patterns from the data using the convolutional layers. 

Second, we adapt the \textit{VGG} network~\cite{VGG} to the one dimensional classification task. We change its number of weight layers to 6 (also test 7, 5, and 4 layers, but find that 6 gives the highest test accuracy of about 99.52\%). However, the drawback is that with fewer convolutional layers, the fully connected layers at the end of \textit{VGG} net become bigger to the point that it hurts the performance (for 4 weight layers it drops to about 95.75\%). This architecture gives us higher accuracy but is rather difficult to adjust to small data.\footnote{The dimensionality of the data is reduced slowly because of the small filter of size 3.}

Finally, one of the strongest and flexible models called \textit{FCN} is based on convolutional neural networks that find general patterns in the time-series sequences~\cite{FCN}. The advantages of the model are: simplicity (no data-specific hyper-parameters), no additional data pre-processing required, no feature crafting required, and significant academic and industrial effort into improving the accuracy of convolutional neural networks~\cite{Band-limiting, Winograd}. 

The architecture of the FCN network contains three blocks, where each of them consists of a convolutional layer, followed by batch normalization $f(x) = \frac{x - \mu}{\sqrt{\sigma^2 + \epsilon}}$ (where $\epsilon$ is a small constant added for numerical stability) and ReLU activation function $y(x) = \max(0, x)$. There are 128, 256, and 128 filter banks in each of the consecutive 3 layer blocks, where the sizes of the filers are: 8, 5, and 3, respectively. We follow the standard convention for Convolutional Neural Networks (CNNs) and refer to the discrete cross-correlation operation as convolution. The input $x$ to the first convolution is the time-series data chunk with a single channel $c$. After its convolution with $f$ filters, the output feature map $y$ has $f$ channels. For training, we insert $s=32$ time-series data chunks into a mini-batch. We have $j \in f$ and the discrete convolution~\cite{FFTconv} that can be expressed as:
\begin{equation}
    y_{(s,j)} = \sum_{i \in c} x_{(s,i)} \star y_{(j,i)} 
\end{equation}

\section{Experimental results}\label{sec:experimental-results}

\subsection{Training and Inference}
Each model is trained for at least 100 epochs. We experiment with different gradient descent optimization algorithms, e.g. Stochastic Gradient Descent (SGD) and Adaptive Moment Estimation (Adam)~\footnote{A very good explanation can be found here: http://bit.ly/2Y9XaQ8}. For the SGD algorithm, we grid search for the best initial learning rate and primarily use 0.0001. The learning rate is reduced on plateau by 2X after 50 consecutive iterations (scheduled patience). SGD is used with momentum value 0.9. We use standard parameters for the Adam optimization algorithm. The batch size is set to $s=32$ to provide high statistical efficiency. The weight decay is set to 0.0001. For our neural network models, the dataset is relatively simple. The \mbox{Wi-Fi} data can be compared in its size and complexity to the MNIST dataset~\cite{mnist} or to the GunPoint series from the UCR archive~\cite{UCRArchive2018}.

\subsection{Time-series width}
\begin{figure}[htb!]
\begin{center}
\includegraphics[width=8cm, height = 5.2cm]{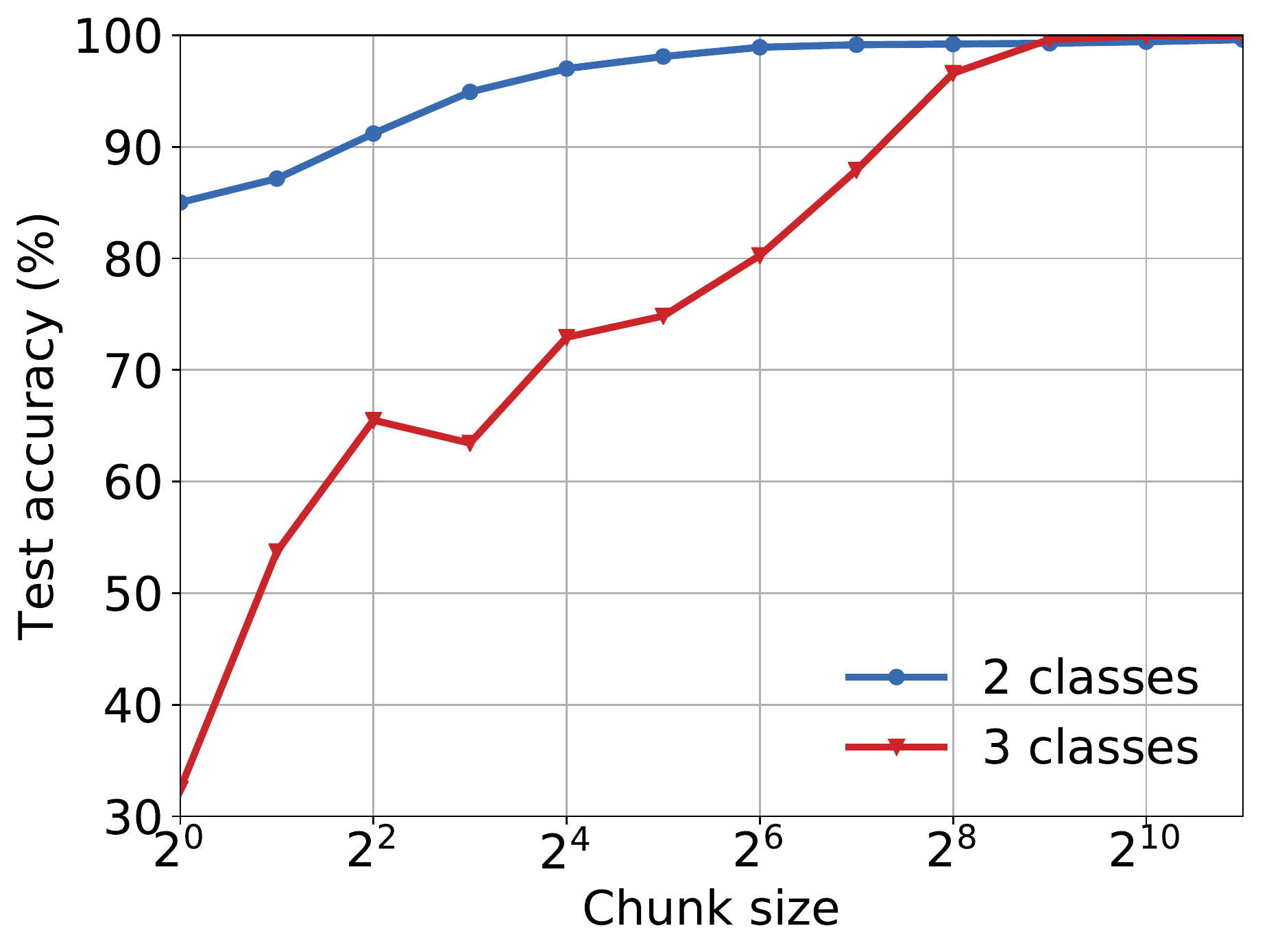}
\caption{The test accuracy (\%) for a model trained and tested for a given chunk size (ranging from 1 to 2048) to distinguish between 2 classes (either 1 or 2 \mbox{Wi-Fis}) and 3 classes (distinguish between 0, 1, or 2 \mbox{Wi-Fis}).}
\label{fig:test-accuracy-chunk-size}
\end{center}
\end{figure}

The number of samples collected per second by the LTE-U BS is about 192. The inference of a neural network is executed in milliseconds and can be further optimized by compressing the network. The final width of the time-series chunk imposes a major bottleneck in terms of the system latency. The smaller the time-series chunk width $w$, the lower latency of the system. However, the neural network has to remain highly accurate despite the small amount of data provided for its inference. Thus, we train many models and systematically vary the chunk width $w$ from 1 to 2048 (see Fig.~\ref{fig:test-accuracy-chunk-size}). In this case, each model is trained only for the single scenario (placement of the \mbox{Wi-Fi} APs) and with zero, one, or two active \mbox{Wi-Fi} APs. When we decrease the chunk sizes to the smaller chunk consisting of a single sample, the test accuracy deteriorates steadily down to the random choice out of the 3 classes (accuracy of about 33\%) and for the 2 classes, its performance is very close to the ED (Energy-based Detection) method.


\begin{figure}[htb!]
\begin{center}
\includegraphics[width=1.0\linewidth]{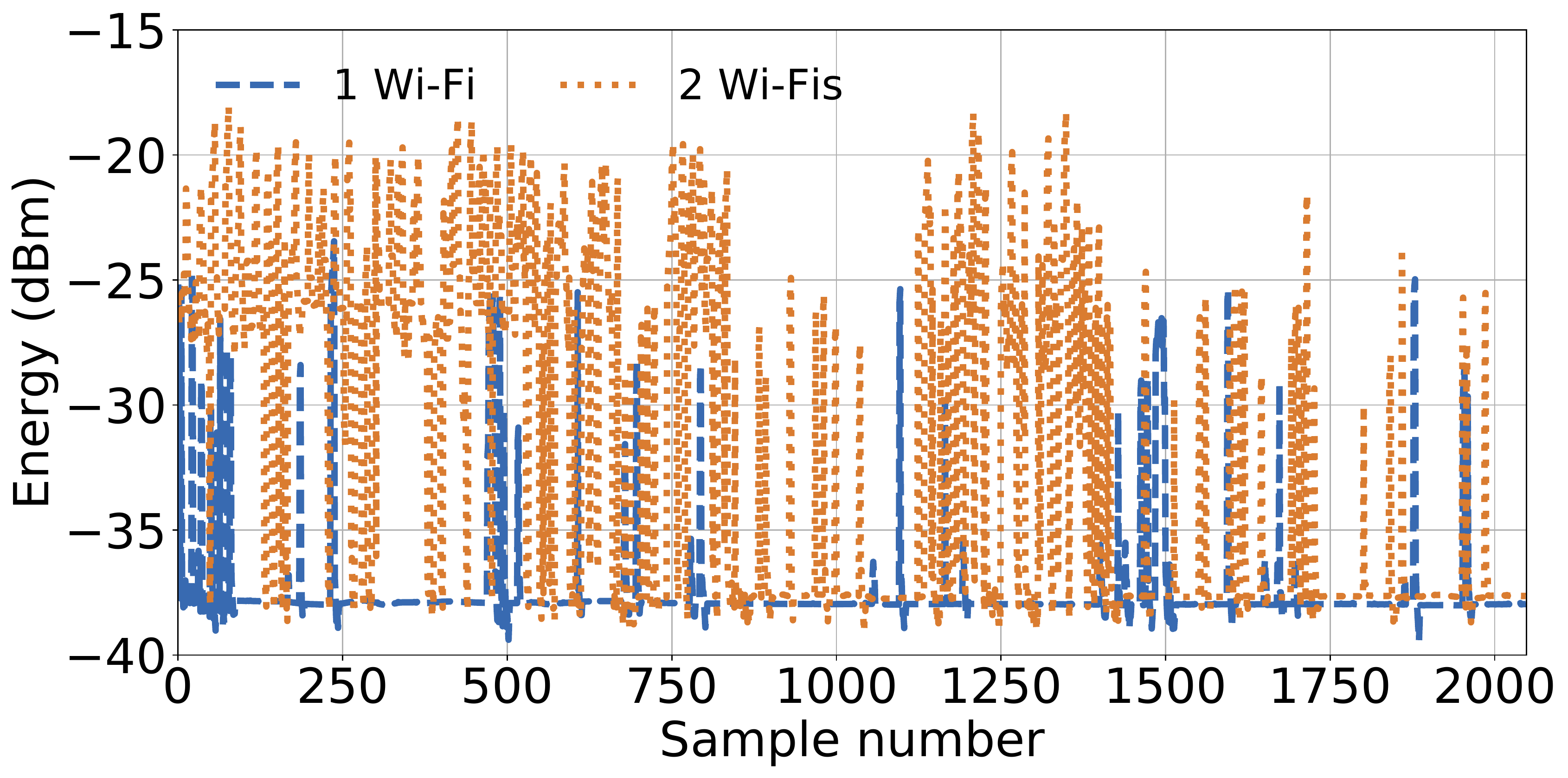}
\caption{The values of the energy (in dBm) captured for 2048 samples in \mbox{LTE-U} BS while there are 1 Wi-Fi, and 2 \mbox{Wi-Fi}s scenarios.}
\label{fig:energy_values}
\end{center}
\end{figure}

In Fig.~\ref{fig:energy_values} we present the values of energy captured for different scenarios with one and two \mbox{Wi-Fi}s. If we consider the signal from about 1500th sample to 2000th sample, it is challenging to distinguish between one or two Wi-Fis. The visual inspection of the signals suggests that width of the time-series chunk should be longer than 500 samples. Signals with width of 384 achieve test accuracy below 99\% and signals with width 512 can be trained to obtain 99.68\% of test accuracy. Based on the experiments in Figs. \ref{fig:test-accuracy-chunk-size} and  \ref{fig:energy_values}, we find that the best trade-off between accuracy and inference time is achieved for chunk of size 512.

\subsection{Transitions between classes}\label{sec:classTransition}

When we switch to another class (change the state of the system in terms of the number of Wi-Fis), we account for the transition period. If in a given window of 1 second a new \mbox{Wi-Fi} is added, the samples from this first second with new \mbox{Wi-Fi} (or without one of the existing Wi-Fis - when it is removed), the chunk is containing values from $n$ and $n+1$ (or $n-1$) number of Wi-Fis. An easy workaround for the \textit{contaminated} chunk is to change the state of the system to new number of Wi-Fis only after the same class is returned by the model in two consecutive inferences (classifications).

\subsection{Real-time inference}

\begin{figure}[htb!]
\begin{center}
\includegraphics[width=8cm, height = 4cm]{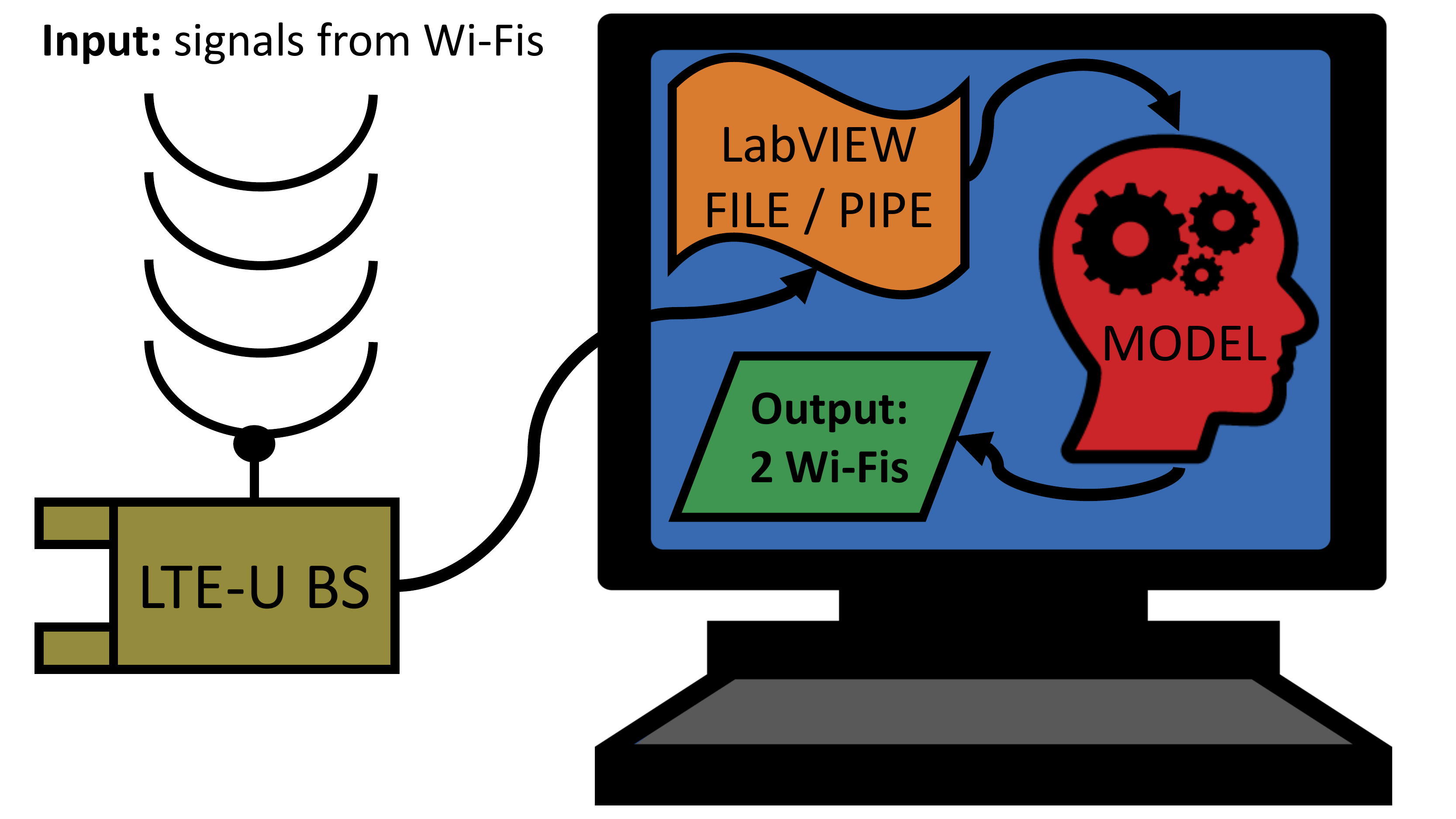}
\caption{The schema of the inference process, where the input received by the \mbox{LTE-U} BS is signals from \mbox{Wi-Fi}s and the output is the predicted number of \mbox{Wi-Fi}s.}
\label{fig:inference}
\end{center}
\end{figure}

We deploy the model in real-time, which similar to the energy data collection experiment setup, and shown in Fig.~\ref{fig:inference}. We prepare the model only for the inference task in the following steps. Python scripts load and deploy the trained PyTorch model. We set up the \mbox{Wi-Fi} devices and generate some network load for each device. The \mbox{LTE-U} BS is connected to a computer with the hardware requirements of at least 8 GB RAM (Installed Memory), 64-bit operating system, x64-based processor, Intel(R) Core i7, CPU clock 2.60GHz. The energy of the \mbox{Wi-Fi} transmission signal in a given moment in time is capture using the NI LabVIEW. From the program, we generate an output file or write the data to a pipe. The ML model reads the new values from the file until it reaches the time-series chunk length. Next, the chunk is normalized and passed through the model that gives a categorical output that indicates the predicted number of \mbox{Wi-Fi}s in the real-time environment.
 
 \section{Performance comparison between ED, AC and ML methods}
 In this section, we analyze and study the performance differences between ED, AC and ML methods  for different configuration setups and discus the inference delay. In ML method, we validate the performance on both test ($ML_t$) and real-time inference ($ML_r$) data. For the final evaluation, we train a single Machine Learning model that is based on the FCN network and used for all the following experiments. The model is trained on the whole dataset of size 3.4 GB, where the train and test sets are of the same size of about 1.7 GB. 

\subsection{Successful Detection at Fixed Distance}
 
\begin{figure}[htb!]
\begin{center}
\includegraphics[width=9cm, height = 5.4cm]{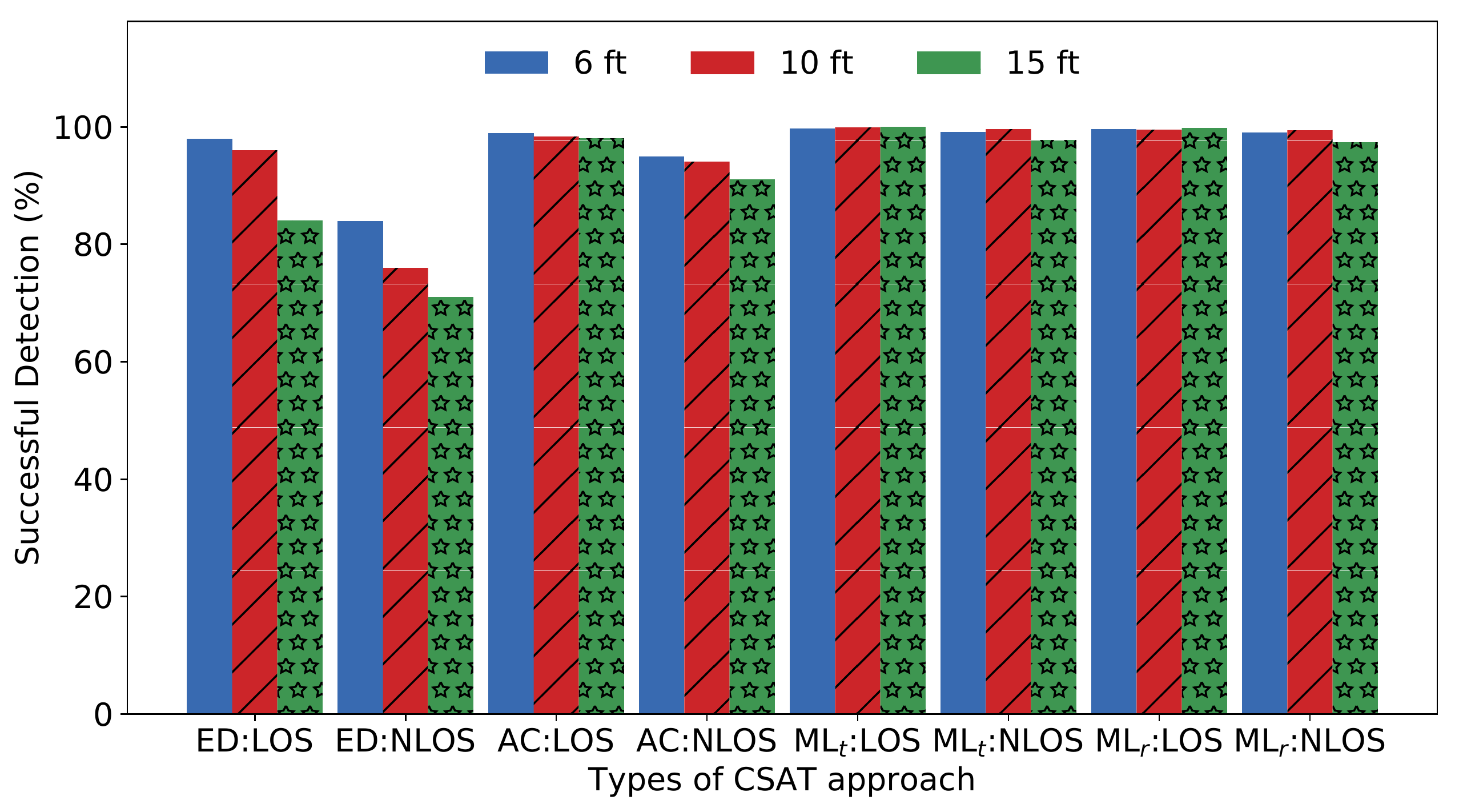}
\caption{Comparison of results for successful detection between ED, AC and ML methods. ML results are presented for the test data (denoted as $ML_t$:) and for the real time inference (denoted as $ML_r$:).}
\label{sample1}
\end{center}
\end{figure}

We compare the $ML_t$ and $ML_r$ performance with ED and AC approaches using the NI USRP platform as shown in Fig.~\ref{exp}. 
In the experiment, \mbox{Wi-Fi} APs are transmitting full
buffer data, along with beacon and probe response frames
following the 802.11 specification. We performed different experiments with 6ft, 10ft and 15ft for LOS and NLOS scenarios.  Fig~\ref{sample1} shows the performance of detection for LOS and NLOS scenarios. In ED and AC based approach the proposed detection algorithm achieves the successful detection on average at 93\% and 95\% for LOS scenario. Similarly, the algorithm achieves 80\% and 90\% for the NLOS scenario. In this work, we show that $ML_t$ and $ML_r$ approach can achieve close to 100\% successful detection rate for both LOS and NLOS, and different distance scenarios (6ft, 10ft \& 15ft). From this result, we observe the $ML_r$ approach works close to the performance of $ML_t$.

\subsection{Successful Detection at Different Configuration}
We verify how the detection works in different configurations. We placed the two \mbox{Wi-Fi} APs on the same side of the LTE-U BS, unlike the above configuration (i.e., 6ft, 10ft and 15ft) where they were on opposite sides. \mbox{Wi-Fi} AP 1 and \mbox{Wi-Fi} AP 2 are placed at distances of 6 feet and 15 feet from the LTE-U BS respectively. We measured the performance of detection with LOS and NLOS configurations.
\begin{itemize}
    \item \textbf{Case A:} Both \mbox{Wi-Fi} AP 1 and \mbox{Wi-Fi} AP 2 are ON.
     \item \textbf{Case B:} Only the \mbox{Wi-Fi} AP 1 at 6 feet is ON.
     \item \textbf{Case C:} Only the \mbox{Wi-Fi} AP 2 at 15 feet is ON.
\end{itemize}
Table~\ref{t1} shows that there is not much degradation in the performance of detection in $ML_t$ and $ML_r$ compared to ED and AC. Hence, we believe that the ML approach is the preferred method for a \mbox{LTE-U} BS to detect the number of \mbox{Wi-Fi} APs and scale back the duty cycle efficiently.
\begin{table}
\centering
\caption{Performance of detection for different configuration setup}
\begin{tabular}{|*{18}{c|}}  
\hline
\multicolumn{1}{|c}{CSAT Types} & \multicolumn{2}{|c}{CASE A (\%)} & \multicolumn{2}{|c}{CASE B (\%)} & \multicolumn{2}{|c|}{CASE C (\%)} \\ \hline 
& LOS & NLOS & LOS & NLOS & LOS & NLOS  \\ \hline
ED & 79 & 76 & 95 & 84 & 93 & 82 \\ \hline
AC & 98 & 94 & 97 & 92 & 96 & 90 \\ \hline
$ML_t$ & 99.96 & 97.74 & 98.80 & 97.96 & 99.94 & 99.37 \\ \hline
$ML_r$ & 99.87 & 97.5 & 98.6 & 97.79 & 99.76 & 99.12 \\ \hline
\end{tabular}
\label{t1}
\end{table}

\subsection{Additional Delay to Detect the \mbox{Wi-Fi} AP}
In ED, the  total  time  for  the  energy  based  CSAT algorithm to adopt or change the duty cycle from 50\% to 33\% is 5.2 seconds (i.e., \mbox{Wi-Fi}  1st  beacon  transmission  time + LTE-U detects $K$ beacon (or) data packets time + NI USRP RIO hardware  processing  time) as shown in Table~\ref{t23}. In AC, the total time for the AC based CSAT algorithm to change the duty cycle from 50\% to 33\% is 4.6 seconds (i.e., \mbox{Wi-Fi} 1st L-STF packet frame + LTE-U detects L-STF frame time + NI USRP RIO hardware processing time). In ML (i.e., $ML_r$), the total time for the CSAT algorithm to adopt the duty cycle from 50\% to 33\% is about 3.0 seconds. This approach is dependent on the chunk size (in this case set to 512). 

\begin{table}
\caption{Other
additional delay to detect the \mbox{Wi-Fi} AP due to the NI hardware 
}
\vspace{-0.2cm}
\centering	
\begin{tabular}{|p{3.5cm}| p{2cm}| }
\hline
CSAT Types & NI HW Delay  \\ 
\hline
Energy Detection (ED) & 5.2 S  \\
\hline
Auto-correlation (AC) & 4.6 S \\
\hline
Machine Learning (ML) & 3 S  \\
\hline
\end{tabular}
\label{t23}
\end{table}

\section{Conclusion}\label{sec:conclusion}
We propose a ML based algorithm that can be used by a LTE-U BS to determine presence of one or two \mbox{Wi-Fi} APs on the channel so that the duty cycle can be adjusted accordingly. We believe that this is the first work to demonstrate the feasibility of using ML on energy values in real-time, instead of packet decoding \cite{Chai:2016:LUS:2973750.2973781}, to reliably distinguish between the presence of different number of \mbox{Wi-Fi} APs. The results show that the ML approach can achieve accuracy close to 100\%  in detection as compared to ED and AC. We aim to extend this work in future by distinguishing between more than two \mbox{Wi-Fi} APs, thus enabling even finer duty cycle adjustments of a LTE-U BS and improved coexistence with \mbox{Wi-Fi}. 

\section*{ACKNOWLEDGEMENT}\label{p4}
This material is based on work supported by the National Science Foundation (NSF) under Grant No. CNS - 1618920. Adam Dziedzic is supported by the Center For Unstoppable Computing (CERES) at the University of Chicago.

\end{document}